\documentclass[5p]{elsarticle}
\usepackage{amsmath,amssymb,amsfonts}
\usepackage{bm}
\usepackage{verbatim}
\usepackage{hyperref}
\usepackage{slashed,braket}
\usepackage{ulem}
\usepackage{cancel}
\usepackage{color}
\usepackage{graphicx,graphics}
\usepackage[title,titletoc]{appendix}
\usepackage{stackengine}
\usepackage{mathrsfs}
\usepackage{tikz}

\newcommand{\tr}{\operatorname{tr}}

\renewcommand{\Im}{\operatorname{Im}}

\newcommand{\ep}{\epsilon}

\newcommand{\D}{\Delta}
\newcommand{\al}{\alpha}

\newcommand{\bp}{\mathbf{p}}

\newcommand{\bk}{\mathbf{k}}

\newcommand{\rev}[1]{{#1}}

\begin{document}

\title{Negative magneto-resistance and Chiral Magnetic Effect in Dirac semimetals 
       from Keldysh technique in Landau levels basis} 

\author{Ruslan A.~Abramchuk}
\ead{ruslanab@ariel.ac.il}
\ead{abramchukrusl@gmail.com}
\address{Physics Department, Ariel University, Ariel 40700, Israel}

\date{\today}

\begin{abstract}
    Negative magneto-resistance, or magnetoconductivity, in Dirac semimetals is conventionally considered as a manifestation of chiral magnetic effect (CME),
    by means of a postulated Chiral Kinetic equation.
    In this paper we study magnetoconductivity in large Fermi energy Dirac semimetals, 
        in one of which (ZrTe$_5$) the effect was observed for the first time. 

    Starting with a Hamiltonian for a quasiparticle model of such a Dirac semimetal,
        we apply the Non-equilibrium Diagram Technique (NDT, or the Keldysh technique)
        to derive the kinetic equations,
        and to investigate the electrons relaxation due to interaction with phonons and disorder in magnetic field.
        Then we calculate the DC magnetoconductivity 
        (the longitudinal to magnetic field component of conductivity)
        as a function of magnetic field strength and temperature. 

    Finally, we compare the obtained temperature dependencies with available to us experimental data.
\end{abstract}

\maketitle

\section{Introduction}
\label{SectIntro}

Magneto-conductivity, or negative magnetoresistance, is one of the characteristic features of Weyl and Dirac semimetals. 
According to the common lore, magneto-conductivity is a manifestation of the chiral magnetic effect. 
The derivation \cite{CMEZrTe5,Kharzeev2017} of relation between these two phenomena is based on the chain of assumptions:
\begin{enumerate}
    \item In parallel electric and magnetic fields the chiral anomaly pumps pairs 
        (left-handed electron and right-handed hole or vice versa) 
        from the Dirac sea of occupied energy levels. 
        Qualitatively, the pumping process results in chiral imbalance 
        (i.e. the difference between the densities $\rho_R$, $\rho_L$ of left-handed and right-handed fermion excitations). 
        The corresponding ad-hoc relaxation-time Chiral Kinetic Equation \cite{CMEZrTe5} for chiral charge density suggests
\begin{equation}
    \rho_5 = \rho_R - \rho_L = N_f\frac{e^2{\vec E} \cdot {\vec B}}{2\pi^2\hbar^2}\,\, \tau_5, \label{EqRho5CME}
\end{equation}
        where $N_f$ is the number of Dirac fermions in the system, 
        and $\tau_5$ is the relaxation time,
            which is deemed to be constant. 

        In this paper, following \cite{CMEZrTe5}, we use SI units, 
        except for absorbing the Boltzmann constant in the definition of temperature 
        in most formulas (\(k_bT^o\text{ [K]}\to T\text{ [meV]}\)).

    \item Chiral chemical potential $\mu_5 =\frac12(\mu_R - \mu_L)$ supposedly exists,
        and is related to the chiral density as if the right- and left-handed fermions are free \cite{CMEZrTe5}
\begin{equation}
    \rho_5 \approx \frac{\mu_5}{3 v_F^3\hbar^3}\Big(T^2 + \frac{\mu^2}{{\pi^2}} \Big), \label{EqMu5CMEw}
\end{equation}
        where $v_F$ is Fermi velocity. 

        In strong magnetic field limit, 
        {if} only the lowest Landau level contributes {to} the dynamics \cite{Kharzeev2017}
\begin{equation}
	\rho_5 \approx \frac{\mu_5}{2 \pi^2 {v_F\hbar^2}} |eB|. \label{EqMu5CME}
\end{equation}

    \item Essentially, CME produces electric current,
         and with the previous assumptions the tensor structure is 
 \begin{equation}
     {\vec j}_\text{CME} = \frac{e^2}{2\pi^2\hbar^2} \mu_5 {\vec B} \label{EqCME}
        \sim \vec B~ (\vec E\cdot\vec B)
 \end{equation}

        In the strong magnetic field the electric conductivity reads \cite{Kharzeev2017}
        $\sigma^{ij}_{CME,B} \sim B^i B^j|B|^{-1}$ and
{\begin{equation}
    \sigma^{zz}_{CME,B} =  N_f \frac{e^2 v_F|e B|}{2\pi^2\hbar^2}  \tau_5 \label{EqCond2CME}
\end{equation}}
        The latter formula was also obtained in \cite{gorbar2014chiral} 
        from the Kubo formula for Weyl fermions,
        without any reference to CME,
\end{enumerate}   

Numerous versions of the calculation 
    that relate the `anomalous' fermionic vector current in Weyl and Dirac semimetals and in QCD 
    to the chiral magnetic effect \cite{Fukushima2008,Kharzeev2017,Stephanov2012ki,Burkov2014,Son2013,Gorbar2016,Gao2012,Hattori2016lqx,Huang2017}. 
Chiral kinetic Theories (CKT) were developed and applied to obtain similar results \cite{Son_2012,Stephanov2012ki,Gao2012,PhysRevB.96.235134,sekine2021axion},
    analyze underlying quantum dynamics \cite{Hattori2016lqx}.
From a more fundamental point of view,

CKT leading to \eqref{EqRho5CME} should emerge from NDT \cite{KamenevBook,Arseev2015}.
A CKT for Weyl semimetal was derived from general principles of QFT in \cite{Lin2019},
    and we adopt some of the techniques in this paper.

Quasiclassical approach \cite{Burkov2014,Son2013,Gorbar2016}, 
    provides concise analysis for systems with \(|\mu|\sim T\) \cite{Burkov2014}.
In $^3$He-A superfluid in which the quasiparticles are electrically neutral,
    a special type of CME is due to emergent gauge field \cite{volovik2017chiral}. 
To the best of our knowledge, all these calculations rely on the three listed assumptions.

The assumption behind \eqref{EqMu5CME} 
    that left- and right-handed fermions exist separately in thermal equilibrium,
    to some extent contradicts the assumption behind \eqref{EqRho5CME}
    that the chiral imbalance is pumped by the chiral anomaly.
Therefore, \eqref{EqMu5CME} describes a stationary state, 
    rather than the true equilibrium. 
Thus the very notion of the chiral chemical potential \(\mu_5\) might contain inconsistency.

Originally, CME \eqref{EqCME} was proposed as electric current along external magnetic field 
    in a medium with chiral imbalance parametrized with the chiral chemical potential. 
Scrutiny revealed that in {the} true equilibrium CME is absent
{\cite{Valgushev:2015pjn,Buividovich:2015ara,Buividovich:2014dha,Buividovich:2013hza,Z2016_1,Z2016_2,nogo,nogo2,BLZ2021}},
see also recent results of numerical lattice simulations \cite{brandt2024absence,buividovich2024out}.

On the other hand, out of equilibrium CME persists, 
    for example \cite{BLZ2022}, for a system with time depending chiral chemical potential.
For the oscillating $\mu_5$ with frequency $\omega$ \eqref{EqCME} is recovered in the limit $\omega \to 0$ if the limit of infinitely large system volume is taken first. 
Therefore, CME \eqref{EqCME} is likely persists in other non-equilibrium setups,
    including the steady state in the presence of external electric field,
    in which case the energy pumped in by the electric field is dissipated during the chirality relaxation. 

Overall, we believe that the three listed assumptions describe the real situation in solids qualitatively correct. 
The lack of rigorous derivation of this pattern, however, 
    leaves a gap in our understanding of the considered phenomena. 
In the present paper we try to close the gap and calculate the magneto-conductivity of Dirac semimetals 
    using kinetic theory obtained directly from NDT (Keldysh technique). 
Our approach does not suffer from the mentioned uncertainties of phenomenological approaches,
    but suggests relaxation time dependence on the external magnetic field and temperature.
We suppose that our calculation holds for the wider class of type-I Weyl semimetals. 

Another fact, which might be seen as a gap in understanding, is
    absolute lack of Landau quantization artifacts in the experimental data
    for the longitudinal conductivity as function of magnetic field
    even in very strong magnetic fields (up to 10-15 T) and very low temperatures.
We discuss the actual observation procedure,
    and suggest a solution for this conundrum in Section \ref{SectBavg}.

Strong magnetic field dependence of the longitudinal conductivity
    demonstrates more complicated pattern 
    than the outlined CME-based understanding.
The raw data on the longitudinal conductivity 
    (see Fig.S1 of \cite{CMEZrTe5}, `Supplemental Materials' in the journal version),
    which is less discussed in the community,
    demonstrates complicated dependence at low temperatures,
    and saturation at intermediate temperatures.
In the original paper \cite{CMEZrTe5} the deviation from the CME-inspired pattern
    was explained by the imperfection of the observation procedure.
    Supposedly, a little misalignment of the magnetic field results in admixture of the transverse (to the magnetic field) conductivity tensor component,
    which essentially decreases with magnetic field strength
.
In our model study the saturation at intermediate temperatures emerges naturally.
At low temperatures the magnetic field dependence is strongly non-monotonous at weak and strong magnetic field limit,
    and neither our nor the original approaches explain this intricate dependence. 

Instead of relying on ad-hoc kinetic equations and other aforementioned assumptions,
    we start with a model Hamiltonian,
    and obtain magneto-conductivity with a sequence of controllable approximations. 
The Hamiltonian contains model parameters, 
    which define fields and coupling strengths.
Aside from those \textit{model} parameters,
    our analysis does not involve \textit{free} parameters.

We consider the model of continuum Dirac fermions as the low energy effective theory 
    that describes physics of Dirac semimetals in vicinity of the Fermi points. 
The electrons interact with disorder and acoustic phonons.
The mechanism for the annihilation of electrons and holes of opposite chirality 
    might be provided, obviously, by small Dirac mass 
    \(mv_F^2 \ll |\mu|\),
    but also by the region of Brillouin zone situated far from the Fermi points,
    where the chiral symmetry is broken by the cutoff. 

Finite results for conductivity 
    without explicitly explaining the chiral symmetry breaking mechanism 
    (the results depend only on the finite level width or relaxation time)
    were obtained in \cite{gorbar2014chiral} (with Kubo formula) 
    and in \cite{Lin2019} (with Chiral Kinetic Theory).
  
The dissipation of energy accompanied by the annihilation of pairs 
    is described by the imaginary part of electron self-energy. 
We demonstrate how this imaginary part appears due to 
    interaction with the thermodynamic ensemble of acoustic phonons and/or scattering on impurities. 

The characteristic features of the semimetals, 
    which simplify our analysis,
    are the relatively small speed of sound (speed of phonons) 
    in comparison to the Fermi velocity for electrons,
    and large chemical potential (Fermi level counted from Fermi points) 
    in comparison to temperatures considered
\begin{equation}
    u\equiv \frac{v_s}{v_F} \ll 1, 
    \quad T \ll |\mu|.
\end{equation}
Thus the initial state for electrons is the degenerate distribution,
    and for phonons -- finite temperature, 
    which enters the final result.

Another simplifying assumption 
    that we use to obtain the simple analytic form of the phononic contributions 
    mixes the previous two
\begin{equation}
    \frac{u}{T}\left(\frac{\hbar \rev{v_F^2}|e B|}{2\pi }\right)^{1/2}, ~ \frac{u|\mu|}{T} \ll 1.
\end{equation}

Impurities and phonons (even soft, because of relatively small sound velocity) 
    provide any momentum transfer, in contrast to QED plasma \cite{Shovkovy2024}.
This unlimited momentum transfer scrambles states at all the LLs crossing the Fermi level.
We start with perturbative NDT treatment in LL basis, 
    then neglect virtual processes,
    and end up with the kinetic theory suggested in \cite{Lin2019}.
Since the virtual processes neglected completely,
    only the imaginary parts of self-energy matter, 
    which we calculate explicitly in Section \ref{SectSE}.
The imaginary parts define the relaxation time.
With the LL basis kinetic theory 
    we calculate the longitudinal to external magnetic field conductivity.
Transverse and Hall conductivities were calculated in \cite{Lin2019} 
    for the strong magnetic field limit, 
    and the corresponding formula for the transverse conductivity 
    can be augmented with our relaxation time,
    while the Hall conductivity is non-dissipative and is independent of the relaxation time.

Though the expression for conductivity \eqref{EqCond2CME} was obtained in various ways, 
    this expression alone  fails to match any experimental data on the dependence of conductivity on magnetic field and temperature  in any parameters range,
    except for the very qualitative notion of magneto-conductivity 
    and qualitative match in weak magnetic field limit at some high enough temperatures.
    --- apparently, the relaxation time parameter $\tau_5$ itself depends on magnetic field and on temperature. 

As in the previous paper devoted to strong magnetic field limit \cite{Abramchuk2024},
    we confirm \eqref{EqCond2CME},  
    but suggest a (model-dependent) dependence of the relaxation time 
    on temperature and magnetic field, with the corresponding half-width \eqref{EqEpsLLL}. 
The obtained result appears to match the experimental data in a certain parameters range, 
    as discussed in Section \ref{SectExp}, Fig.\ref{FigRhoLLLT}

For arbitrary magnetic field,
    our resulting formula \eqref{EqCondIIExt} is well-defined and reproduces the zero and strong magnetic field limits.
Our formula involves explicit summation via LLs.
The invisibility of LLs in experimental data is discussed in Section \ref{SectBavg}.
In the weak magnetic field the formula reproduces the quadratic magneto-conductivity pattern of \eqref{EqCME}.

The paper is organized as follows.
In the rest of this section we discuss application of the Drude formula to the zero and strong magnetic field cases.
In Section \ref{SectKeK} we derive gauge-invariant kinetic equations for the electrons from NDT.
In Section \ref{SectMC} we obtain the main result --- longitudinal conductivity in magnetic field.
In Section \ref{SectSE} we calculate the electrons level widths (dissipation rates),
    which define the relaxation time,
    from the electrons self-energy imaginary part.
In Section \ref{SectExp} we try to compare our theoretic results with available experimental data.
Section \ref{SectCon} is for brief results summary and conclusions.

\subsection{Conductivity from Drude formula}

Longitudinal conductivity is classical in a sense and thus is described by the Drude formula.
The Drude formula for conductivity explains Ohmic conductivity, 
    which is finite because of momentum scattering.
Though the Drude formula holds in NDT \cite{KamenevBook}, Eq. (11.55) 
    the derivation is highly non-trivial and goes beyond perturbation theory in coupling constant, 
    and also require methods of kinetic theory, which probably is equivalent to summation of certain subseries.
The formula for DC case reads
\begin{gather}
    \sigma = e^2\nu D, \quad D = \frac{v_F^2\tau_\text{tr}}{d},
\end{gather}
where \(e\) is the carrier (typically, the electron) charge, 
    \(\nu\) \rev{spectral} density of its states,
    \(v_F\) its velocity.
\(\tau_\text{tr}\) is the relaxation time,
    and \(d\) is the number of spatial dimensions.
The combination \(D\) is the diffusion coefficient, 
    which describes diffusion of the carriers momentum in the diffusion approximation.
The relaxation time can be figured out from the underlying QFT for the medium.
In the leading order the relaxation time can be estimated from the perturbation theory as an imaginary renormalization of the carrier's energy, 
    the half-width or the dissipation rate, denoted as \(\ep=\frac{\hbar}{2\tau_\text{tr}}\), 
    which we calculate with NDT in Section \ref{SectSE}.

Without the magnetic field, 
    the velocity is again the Fermi velocity,
    \(d=3\), 
    density of states on the Fermi sphere surface with the leading factor of 2 due to the spin degeneracy
    \(\nu = 2\frac{4\pi\mu^2}{(2\pi\hbar)^3v_F^{\rev{3}}}\),
    and \(\tau_\text{tr} = \frac{\hbar}{2\ep_\mu}\) with \(\ep_\mu\) given by \eqref{EqEpsMu} 
\begin{align}
    \sigma_0 &= e^2~2\frac{4\pi\mu^2}{(2\pi\hbar)^3\rev{v_F^3}}~\frac{v_F^2}{3}~\frac{\hbar}{2\ep_\mu} 
        = \frac{e^2\mu^2}{6\pi^2\hbar^2\rev{v_F}\ep_\mu} \nonumber\\
    &= \frac13(\rho_\text{imp}+\rho'_\text{ph}T)^{-1} \label{EqCond0}
\end{align}
Let's also track how the spatial factor of \(\frac1d, ~d=3\) emerges.
Assuming the conducting states are the states at the Fermi surface \(E=\mu\), 
    with the spectrum \(E=v_F|\vec p|\), and velocity 
    \(v_i = \frac{\partial E}{\partial p_i}=\frac{v_F^2 p_i}{E}\),
    and we omit the normalization factors  since they cancel out, 
    as well as the integration over transverse degrees of freedom in 3D with the cylindrical coordinates
\begin{gather}
    \braket{v_z^2} = \frac{v_F^4}{\mu^2}\frac{\int dp_z ~p_z^2 \Theta(\mu^2-v_F^2p_z^2)}{\int dp_z ~ \Theta(\mu^2-v_F^2p_z^2)} = \frac{v_F^2}{3}.
\end{gather}

\subsection{Drude formula for Strong magnetic field }

Application of the Drude formula to the strong magnetic field limit, 
    in which the LLs splitting is larger than the chemical potential \(2|eB|\hbar v_F^2>\mu^2\),
    is straightforward
For simplicity, let us disregard the small gap at the Fermi point in the Dirac spectra,
    than the LLL spectra is linear \(E_\text{LLL}=v_F|p_z|\).

The parameters of the Drude formula are readily written.
The velocity is the Fermi velocity in the Dirac equation.
The density of states \(\nu = 2\frac{|eB|}{(2\pi)^2}\) comes from the magnetic flux quanta \cite{LL3},
    divided by additional \(2\pi\) for \(p_z\) density of states,
    and with the leading factor of 2 accounting for the {\itshape two} points where the Dirac equation (as opposed to Weyl) spectra crosses the Fermi level.
    The spin degeneracy is lifted by magnetic field at LLL.
    \(d=1\) since the motion is effectively one-dimensional (the `ultra-quantum' regime).
And finally, with the relaxation time \(\tau_\text{tr}=\frac{\hbar}{2\ep_\text{LLL}}\) 
    we again obtain the result of  \cite{gorbar2014chiral,Kharzeev2017} (obtained for some phenomenological relaxation parameter)
\begin{gather}
    \sigma_\text{LLL} = \frac{|eB|e^2v_F}{4\pi^2\hbar\ep_\text{LLL}}  \label{EqCondLLLt}
\end{gather}
which reads with our \(\ep_\text{LLL}\) 
    determined by \eqref{EqEpsLLL} \(\ep_\text{LLL} \sim |eB|\),
\begin{gather}
    \sigma_\text{LLL} = ({\rho_\text{imp}+\rho'_\text{ph}T})^{-1}. \label{EqCondLLL}
\end{gather}
as long as the magnetic field satisfies \eqref{EqEpsLLL}.
The result suggest plateau as a function of magnetic field,
    and, apparently, matches \cite{Abramchuk2024} the experimental data for ZrTe$_5$ \cite{CMEZrTe5} 
    as a function of temperature $T$.

As expected, the system exhibits magneto-conductivity \(\sigma_\text{LLL}=3\sigma_0>\sigma_0\).
The enhancement of conductivity by the factor of 3 is apparently due to the dimensional reduction in the strong magnetic field.

However, application of the Drude formula to the case of a finite magnetic field is not so obvious.

\section{Kinetic equations from the Keldysh technique}
\label{SectKeK}

To reach general understanding of how kinetic equations emerge from a QFT we follow
    \cite{KamenevBook} (in particular, Eqs. (9.67), (11.4)-(11.9)) and Sect. 5 of \cite{Arseev2015}.
To generalize the analysis to the case of external magnetic field we then follow \cite{Lin2019}.

\subsection{Kinetic equations from QFTs}

``Kinetic equations'' emerge from the lesser component of the Dyson equation.
In the triangular representation with the lesser component in the upper right corner 
\begin{gather}
    \hat{\mathbf{G}} = \begin{pmatrix}
        \mathbf{G}^R & \mathbf{G}^< \\ & \mathbf{G}^A
    \end{pmatrix}, 
\end{gather}
the equation reads,
    where all the neighboring symbols are convoluted rather than multiplied
\begin{gather}
    (\hat Q-\hat{\mathbf{\Sigma}})\hat{\mathbf{G}} = \hat\delta \iff
    \begin{cases}
        (Q^{R/A}-\mathbf{\Sigma}^{R/A})\mathbf{G}^{R/A}=\delta \\
        (Q^R-\mathbf{\Sigma}^R)\mathbf{G}^< = \mathbf{\Sigma}^<\mathbf{G}^A
    \end{cases}
\end{gather}
The bold symbols stand for the exact Green functions and self-energy parts, 
    while the regular symbols are for the free Green functions and equation of motion operators
\( 
    Q={G^{R}}^{-1}, Q^* = \overleftarrow{{G^{A}}^{-1}}. 
\) 
We generally assume that the exact Green functions and self-energy parts are conjugated as their free counterparts
\begin{gather}
    G^{R/A\dag}=G^{A/R}, ~G^{<\dag}=-G^<, \\
    \Sigma^{R/A\dag}=\Sigma^{A/R}, ~\Sigma^{<\dag}=-\Sigma^<. \nonumber
\end{gather}

In absence of external gauge fields in non-relativistic systems, 
the shortest path \cite{KamenevBook,Arseev2015} to Fokker-Plank type equations is to 
utilize the parametrization 
    which automatically ensures the anti-hermiticity of the lesser Green function
\begin{gather}
    \mathbf{G}^< = \mathbf{nG}^A - \mathbf{G}^R\mathbf{n}, 
        ~\mathbf{n}^\dag = \mathbf{n}, \label{EqGLn}
\end{gather}
The lesser Dyson equation multiplied by \((Q-\mathbf{\Sigma}^R)^\dag\) from the right
than reduces to 
\(
    Qn-nQ^* = \mathbf{\Sigma}^< + \mathbf{\Sigma}^Rn -n\mathbf{\Sigma}^A \label{EqDysonKin},
\)
which in the quasi-classical limit with all the virtual processes disregarded
    produces standard kinetic equations.
The left hand side reduces to the kinetic terms, and the right --- to the collision integral.
The LHS at \(t=t'\) is local in time, 
    while the RHS is essentially non-local because of its quantum nature.
The self-energy parts can be specified with the perturbation theory for a given QFT.
Standard collision integrals emerge from RHS if virtual processes are neglected.
However, such \(n(x,y)\) is not gauge invariant in general, 
    and thus we adopt another approach.

\subsection{Kinetic equation for Relativistic system in external Gauge fields}

In presence of the emergent relativistic symmetry and in external gauge fields
\begin{gather}
    Q = i(\partial_t -ieA_0(x)) -v_F\al^ki(\partial_{x_k} -ieA_k(x)) \label{EqRelQ}
\end{gather}
    we generally follow \cite{Lin2019}.
The central object \(W(x,p)\) in this approach is 
    the lesser component Green function augmented with the parallel transporter, 
    which makes \(W\) manifestly gauge invariant.
In this paper we study the electric current
\begin{gather}
    j_\mu(x) = \tr\gamma_0\gamma_\mu iG^<(x,x) = \tr\gamma_0\gamma_\mu\int\frac{d^4p}{(2\pi)^4}W(x,p) \\
    iW(x,p) = \int d^4y e^{ipy} G^<\Big(x+\frac y2,x- \frac y2\Big) U_{x-\frac y2,x+\frac y2}\\
    U_{xx'} = \exp\left(-ie\int_{x'}^{x} A_\mu(z)dz^\mu\right)
\end{gather}
The equations for \(W(x,p)\) are the Wigner-transform of the Dyson equations multiplied by the parallel transporter \cite{Xavier2024}.
The multiplication is trivial since the parallel transporter passes the covariant derivative by definition.
The Wigner transform in case of uniform external electromagnetic field is exactly calculable,
    as we seen in \cite{Abramchuk2024} for external magnetic field,
    because the integral is Gaussian 
    since the gauge can be fixed exactly (e.g.~the Fock-Schwinger gauge) and
\begin{gather}
    A^\mu(x) = \frac12x^\nu F_{\mu\nu} \\
    U_{xy} = \exp\left(ieF_{\mu\nu}\int_x^y\frac{z^\mu}{2}dz^\nu\right) 
    = \exp\big(ieF_{\mu\nu}x^\mu y^\nu\big)
\end{gather}

The advantage of the manifestly gauge invariant definition is that,
    if the chiral invariance persists at the Fermi surface, \([Q,\gamma_5]=0\),
    the hermitian \(W\) 
    can be decomposed as
\begin{gather}
    W(x,p) = \frac14 (\rho(x,p) + \vec\al\cdot\vec j(x,p)) \\
    J_0(x) = \int\frac{d^4p}{(2\pi)^4}\tr W(x,p) = \int\frac{d^4p}{(2\pi)^4}\rho(x,p) \\
    J_i(x) = \int\frac{d^4p}{(2\pi)^4}\tr\al_i W(x,p) = \int\frac{d^4p}{(2\pi)^4}j_i(x,p),
\end{gather}
and the equations for \(W\) and thus for \(\rho,j_i\) are gauge invariant.

After the Wigner transform, the covariant derivatives turn into \cite{Vasak1987}
\begin{gather}
    \D_\mu = \partial_{x^\mu} - F_{\mu\nu}\partial_{p^\nu} \\
    \D_0 = \partial_t - E^k\partial_{p_k},
    \D_i = \partial_i + \ep^{ilk}B_k\partial_{p_l} \nonumber
\end{gather}
in case of the uniform external electric and magnetic fields.

The transforms of the Dyson equation for the lesser component and its conjugate then read
\begin{align}
    \Big(\frac12\D_\mu &-ip_\mu\Big) \gamma_0\gamma^\mu W(x,p) = \label{EqDW}\\ 
    &= W[\int_z(\mathbf{\Sigma}^<_{x_1z}\mathbf{G}^A_{zx_2}+
        \mathbf{\Sigma}^R_{x_1z}\mathbf{G}^<_{zx_2})U_{x_1x_2}] \nonumber\\
    \Big(\frac12\D_\mu &+ip_\mu\Big)W(x,p) \gamma_0\gamma^\mu = \nonumber\\
    &=-W[\int_z(\mathbf{G}^R_{x_1z}\mathbf{\Sigma}^<_{zx_2}+
        \mathbf{G}^<_{x_1z}\mathbf{\Sigma}^A_{zx_2})U_{x_1x_2}] \nonumber
\end{align}
Essentially, to obtain the simple relaxation time approximation, 
    we disregard all the virtual processes.
In special cases, a purely virtual processes can provide finite Drude weight,
    as was recently reported for flat-band materials at two-loop level \cite{Holder2024}
    --- we do not expect such effects.
The one-loop perturbative self-energy at the Fermi level provides the `collision-related' RHS ,
    with the parametrization \eqref{EqGLn}
    and renormalization terms omitted
\begin{gather}
    \mathbf{\Sigma}^<_{12} \approx \Sigma^{(1)<} \approx 2i\hbar^{-1}\ep \mathbf{n}_{12}, \quad
        \mathbf{\Sigma}^{A/R}_{12} \approx \pm i\hbar^{-1}\ep \delta_{12},  \label{EqSigmaKin}\\
    \mathbf{\Sigma}^<\mathbf{G}^A-\mathbf{G}^R\mathbf{\Sigma}^< 
        + \mathbf{\Sigma}^R\mathbf{G}^< - \mathbf{G}^<\mathbf{\Sigma}^A
        \approx {2}i\ep\mathbf{G}^<  \nonumber
\end{gather}
where (see \eqref{EqSigmaLL},\eqref{EqEpsMuB} of Section \ref{SectSE})
\begin{gather}
    \ep(B)=\sum_{n:E_{n,p_z=0}<\mu}\ep_n.
\end{gather}
The sum is over the LLs that cross the Fermi surface.

In contrast to our case, in \cite{Shovkovy2024,Ghosh2024,Wang2021} 
    the limit \(T\gg\mu\) was considered.
The fermions 
    interact with a thermal ensemble of photons, 
    so the transitions between energy levels with negligible momentum transfer is essential.
We consider the opposite limit \(\mu\gg T\), 
    disregarding virtual processes whatsoever,
    since the disorder is dominant and momentum transfer is unlimited.


The dynamic equations are the trace of the sum of the Wigner-transformed Dyson equations \eqref{EqDW} in the relaxation time approximation
\begin{gather}
    \begin{cases}
        \D_0\rho + \D_ij_i  \approx -{2}\ep \delta\rho \\
        \D_0j_i + \D_i\rho - 2\ep^{ilk}p_lj_k \approx -{2}\ep \delta j_i 
    \end{cases}\label{EqLLKin}
\end{gather}
where the half-width \(\ep = \ep(B,T,\mu)\) is estimated in \eqref{EqEpsMuB}
and \(\delta\rho, \delta j_i\) measures the deviation of \(\rho,j_i\) from the equilibrium values \(\tilde \rho,\tilde j_i\).
The difference of the equations \eqref{EqDW} produces constraints \cite{Lin2019}. 

\section{Longitudinal Magneto-conductivity from the kinetic equations}
\label{SectMC}

The system \eqref{EqLLKin} was solved in \cite{Lin2019} to find transverse and Hall conductivities
    in the strong magnetic field limit
(the case of transverse electric and magnetic field \(\vec E\bot\vec B\)). 
For the conductivities \(\sigma_\bot\) and \(\sigma_H\)  in the strong magnetic field limit see Eqs. (28-29) in Section III of \cite{Lin2019}, correspondingly.
The Hall conductivity exemplifies non-dissipative transport, 
    and thus is well-defined even for infinite relaxation time.
We suggest that the relaxation time in the expression for \(\sigma_\bot\) is defined by our \eqref{EqEpsLLL} as \(\tau=\frac12\hbar\ep^{-1}(\mu,B)\).

For parallel electric and magnetic fields \(\vec E\parallel\vec B\), 
    the static linear response to the electric field reads 
\begin{gather}
    -E_\parallel \partial_{p_z}\tilde j_\parallel = -\ep(B)\delta j_\parallel \\
    J_\parallel = \sigma_\parallel E_\parallel = \int\frac{d^4p}{(2\pi)^4}\delta\tilde j_\parallel \nonumber\\
    \sigma_\parallel = \frac{1}{\ep(B)}\int\frac{d^4p}{(2\pi)^4}\partial_{p_z}\tilde j_\parallel
\end{gather}
where \(\tilde j_\parallel(x,p)\) is the equilibrium distribution in the magnetic field,
    which can be read of the exact spectral function in the magnetic field, see Eq.(12) of  \cite{gorbar2014chiral} with magnetic field \(l\)
\begin{gather}
    \tilde j_\parallel(x,p) = 2\pi e^{-p_\bot^2l^2}\sum_{n\ge 0}(-1)^n(-2v_Fp_z)
        (L_n(2p_\bot^2l^2)-\nonumber\\
    -L_{n-1}(2p_\bot^2l^2)) \delta(\mu^2-E_{np_z}^2) f(\mu-p_0) 
\end{gather}
with the Fermi distribution \(f\), and the relativistic Landau spectra
\begin{equation}
    E_{np_z}^2=v_F^2(p_z^2 + 2|eB|\hbar n),
    \quad n\ge 0.
\end{equation}
In contrast to \cite{Lin2019}, we consider \(\mu\gg T\) 
    so the terms containing \(\partial_{p_0}f\) are dominant.

With the transverse momentum integration as in \eqref{EqSumPtn}
    we in fact reproduce the Drude formula 
(during the integrations over \(p_0,p_3\) we drop terms 
    that are odd in momentums and thus eventually vanish)
\begin{align}
    \sigma_\parallel &= \frac{e^2}{\ep(B)}\sum_{n\ge 0}(2-\delta_{n0})
        \int\frac{dp_0dp_z |eB|}{(2\pi)^2} \frac{p_z^2}{E_p^2}\times\nonumber \\
    &~\times \delta'(p_0-E_p)f(\mu-p_0) \nonumber\\
    = &\frac{e^2v_F^4}{\ep(B)}\sum_{n\ge 0}(2-\delta_{n0}) \frac{|eB|}{2\pi} 
        \int\frac{dp_z}{2\pi} \frac{p_z^2}{\mu^2}(-f'(\mu-E_{np_z})) \nonumber\\
    &\approx \frac{e^2v_F}{2\hbar\ep(B)}\frac{|eB|}{2\pi^2} \sum_{n= 0}^{N_B}(2-\delta_{n0}) 
        \sqrt{1-\frac{n}{\nu_B}},  
    \label{EqCondII}
\end{align}
where the distribution is degenerate \(-f'(\mu-E)\approx\delta(\mu-E)\),
    and thus the number of the highest Landau level that cross the Fermi level is 
    \(N_B = \left\lfloor \nu_B \right\rfloor, ~\nu_B=\frac{\mu^2}{2|eB|\hbar v_F^2}\).

For any magnetic field the series over the LLs can be computed numerically 
    with the dissipation provided by \eqref{EqEpsMuB},\eqref{EqRhos} and the result is manifestly finite
    (though non-monotonous)
\begin{gather}
    \sigma_\parallel = \frac{ \sum_{n= 0}^{N_B}(2-\delta_{n0}) 
        \sqrt{1-\frac{n}{\nu_B}} }
    {\big(\rho'_\text{ph}T+\rho_\text{imp}\big) 
        \sum_{n'=0}^{N_B}\frac{(2-\delta_{n'0})}{\sqrt{1-\frac{n'}{\nu_B}}}
    } \label{EqCondIIExt}  
\end{gather}
The expression demonstrates oscillatory behavior
    that we discuss at the end of this section, see \eqref{EqCondG}, Fig.\ref{FigRhoG}.

The formula interpolates between the strong magnetic field \eqref{EqCondLLL} (with only \(n=n'=0\) terms non-zero) 
    and vanishing magnetic field \eqref{EqCond0} limits, since 
\[\sum_n\to\int_0^1dx\sqrt{1-x}=\frac23, ~\sum_{n'}\to\int_0^1\frac{dx'}{\sqrt{1-x'}}=2\]
The spatial factor of \(\frac13\) emerged naturally from the  sum over a  large number of the LLs. 

The weak field magnetoconductivity can be addressed analytically further.
Let us calculate the sums over LLs in \eqref{EqCondIIExt}
    with the Euler-Maclaurin formula in the leading order
    (Eq. (59.10a) in \cite{LL5}) 
\begin{align}
    \sum_a^b f(n) &\approx \int_a^b f(x) dx +\\
    &+ \frac{f(a)+f(b)}{2}  + \frac{f'(b)-f'(a)}{12}.\nonumber
\end{align}
Aside from the regular terms in powers of \(\nu^{-1,-2}\)
    we encounter `oscillating' terms proportional to 
    \((\nu-\lfloor\nu\rfloor)^{+\frac12,-\frac12,-\frac32}\)
    in numerator and denominator.

If we just disregard the oscillating terms (see the next subsection), 
    the weak field conductivity reads 
\begin{align}
    \bar\sigma_\parallel^{(2)}(B,T) &\approx 
        \frac{\sigma_\text{LLL}}{3}\Big(1 + \frac{\nu^{-2}}{6}\Big) \nonumber\\
    &= \sigma_0(T)\Big(1 
        + B^2\frac{2 e^2\hbar^2v_F^4}{3\mu^4}
    \Big)
    \label{EqCondWF}
\end{align}
The zero order reproduces \eqref{EqCond0} 
    --- the HLLs weights add up precisely to reproduce the dimensional factor.
The linear term vanishes, though it might reappear, 
    if we more thoroughly consider the summation.
The positive quadratic magneto-conductivity emerges specifically because of the relativistic dispersion,
    and is similar to the CME-based \cite{CMEZrTe5} magnetoconductivity term as function of magnetic field, chemical potential and relaxation time \(\sim B^2\mu^{-2}\tau\), c.f. \eqref{EqMu5CMEw},\eqref{EqCME}.

\subsection{Invisibility of the Landau quantization}
\label{SectBavg}

Let us now discuss the oscillations problem.
\rev{The series over LLs instead of integrals over energy emerged 
    because we disregarded electrons temperature.
Eventually, at temperatures larger then LLs energy splitting,
    the oscillations disappear.
However, in the ``large \(\mu\)'' materials, which we discuss in this paper,
    temperatures of order of several or dozens of K are not enough to smear the oscillation completely.
Thus we suggest an alternative mechanism which might explain 
    the absence of the Landau quantization artifacts even at several K temperatures in experimental data.}

In any realistic setup, the value of magnetic field varies as a function of the transverse coordinate 
    \(b=B_\parallel(\vec x_\bot)\) 
    with an average \(B=\braket{b}_b\) 
    and a finite variance \(\delta B=\sqrt{\braket{(b-B)^2}_b}\).
As a prominent example, the renowned spin echo effect is determined by 
    this microscopic stochasticity of a macroscopically uniform magnetic field \cite{VanKampen1976}.

\begin{figure}
	\includegraphics[width=0.9\linewidth]{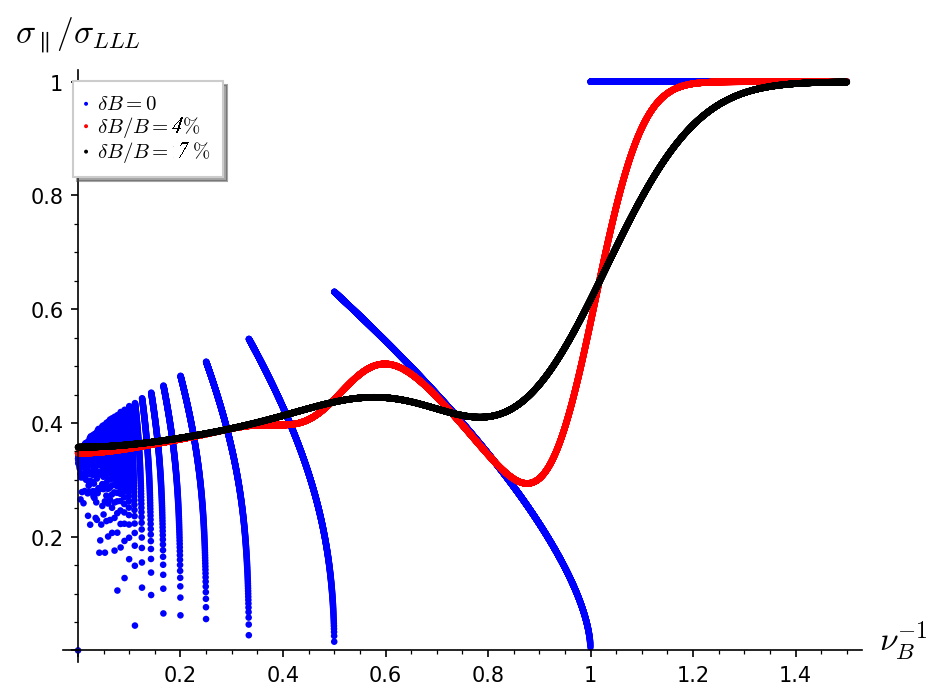}
	\includegraphics[width=0.9\linewidth]{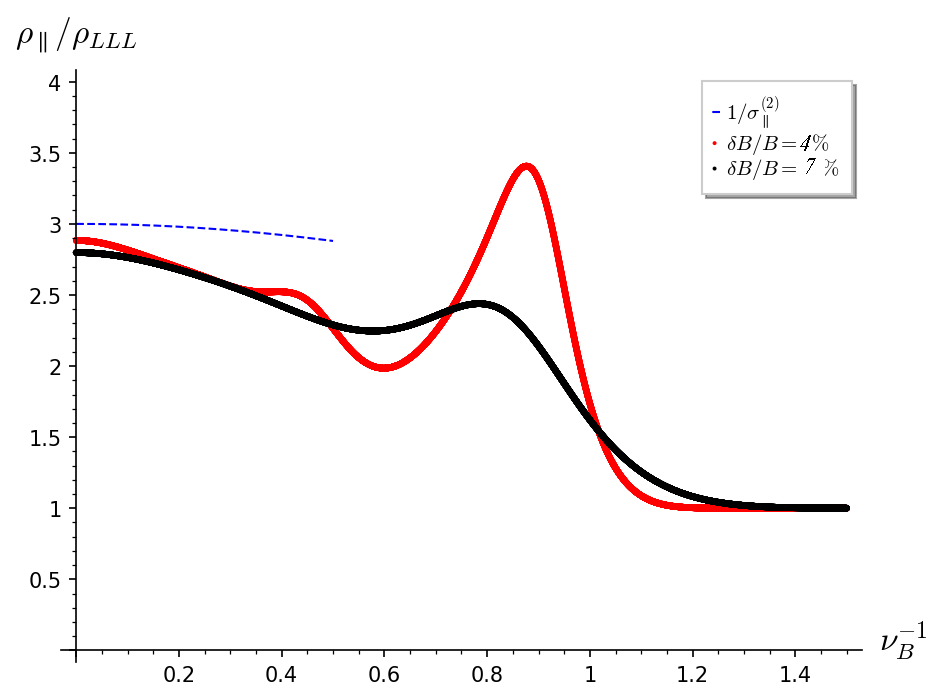}
    \caption{Averaged longitudinal conductivity \(\bar\sigma_\parallel\) \eqref{EqCondG}
        and resistivity \(\rho_\parallel=\bar\sigma_\parallel^{-1}\) 
        as functions of magnetic field.
    The blue dots are for non-smoothened \(\sigma_\parallel\) \eqref{EqCondIIExt}.
    The red line (stronger oscillating) corresponds to \(\delta B/B=0.0\rev{4}\) and
    the black line (weaker oscillating) --- to \(\delta B/B=0.\rev{07}\).
    The dashed line --- the quadratic magneto-conductivity \eqref{EqCondWF}.
    \(\nu_B^{-1}=\frac{2|eB|\hbar v_F^2}{\mu^2}\) so that \(\lfloor \nu_B\rfloor = N_B\)
        is the number of the highest occupied LL (LLL's number is zero).
        \label{FigRhoG}
	}
\end{figure}

To model the varying magnetic field 
    we convolute the `discrete' formula \eqref{EqCondIIExt} with a Gaussian kernel
\begin{align}
    \bar\sigma_\parallel(B,T) &= \braket{\sigma(b,T)}_b\nonumber\\
    &= \int \frac{db }{\sqrt{\rev{2}\pi}\delta B} \exp\Big(-\frac{(B-b)^2}{\rev{2}(\delta B)^2}\Big) 
        \sigma_\parallel(b,T)\label{EqCondG}
\end{align}
A smooth function passes the convolution, 
    while oscillations get filtered.
Thus we expect that the analytic formula \eqref{EqCondWF} reproduces \eqref{EqCondG} 
    for some large enough variance at weak enough magnetic field. 
For an intermediate magnetic field strength, we couldn't calculate the convolution analytically,
    for the numerical analysis see Fig.\ref{FigRhoG}.
For \(\delta B\to+0\) as expected \(\bar\sigma_\parallel/\sigma_0\to 1-0\) at \(B=0\).
The dependence for smaller magnetic field variance oscillates stronger (the red line),
    than for larger variance (black line).

The conductivity is to be averaged, rather than resistivity
    since each thin section of the material along the magnetic field conducts in parallel.
On the experimental side, resistance \(R_\parallel\) is measured
    and the resistivity is proportional for a cylindrical sample
\begin{gather}
    \rho_\parallel(B,T) = \frac{R_\parallel(B,T) S_\bot}{L_\parallel} 
        = \bar\sigma^{-1}_{\parallel}(B,T)
        = \braket{\sigma(b,T)}_b^{-1}
\end{gather}
Variance of order of few percent 
    smothers the LL quantization, 
    and turns \(\sigma_\parallel\) \eqref{EqCondIIExt} into something similar to the experimental data,
    see Fig.\ref{FigRhoG}.

\rev{Also, Material inhomogeneity might enhance the effective magnetic field variance, 
    and the finite temperature effect is always present in practice. }

\section{Electrons self-energy}\label{SectSE}

In this section we calculate the relaxation time in a simplistic model with disorder and acoustic phonons. 
The relaxation time is the inverse of the energy band width at the Fermi level,
    and the half-width is provided by the imaginary part of the self energy.
The real part of self-energy was studied in the strong magnetic field limit case in \cite{Abramchuk2024}.

The NDT one-loop correction reads
\(
    \hat G_1 = \hat G\hat \Sigma\hat G,
\)
where \cite{Arseev2015}
\begin{gather}
    \Sigma^<=ig^2G^<D^<,  \\
    \Sigma^R = ig^2(G^RD^<+(G^<+G^R)D^R),\nonumber
\end{gather}
where the loop momentum integrals are implied for a two-point Green function \(D\) 
    (we consider disorder and thermal ensemble of phonons)
\begin{gather*}
    \Sigma=ig^2 GD\equiv\Sigma_p=ig^2\int_kG_{p-k}D_{k}, ~\int_k\equiv\int\frac{d^4k}{(2\pi)^4}.
\end{gather*}
The free lesser component of the Keldysh-Green function for a trivial distribution reads
\begin{gather}
    G^< = (G^A-G^R)f(p_0).
\end{gather}
The Dyson equation allows to sum up diagram subseries 
\begin{gather}
    \hat Q_{(1)} = \hat Q - \hat\Sigma,
    \quad \hat G_{(1)} = (\hat Q - \hat \Sigma)^{-1}, \label{EqDyson}
\end{gather}

From \eqref{EqDyson} we find self-consistency relations for the half-width,
which is approximately equal to the self energy anti-hermitian part
(the imaginary scalar part of the one-loop self-energy due to the electron-phonon interaction and disorder).
The remaining terms yield the self-energy renormalization 
\(\Sigma^R_\text{(ren)} = ig^2(G^R+G^<)D^R\),
since they are not coupled to the ensemble of phonons, 
    and thus cannot contribute to energy dissipation.

In a sufficiently strong magnetic field,
    the half-width \(\ep\) itself depends on the magnetic field,
    while in the weak magnetic field the dependence should be neglected.

The non-zero self-energy itself yields transitions between the LLs \cite{Shovkovy2024}
Thus the ansatz for solving the kinetic equations \eqref{EqLLKin} 
    that the distribution \eqref{EqGLn} depends only on energy \(\mathbf{n}\to f(E)\)
    is justified.
See \cite{Ghosh2024} for LL Green's functions; 
    \(G = U\bar G)\), where \(\bar G\) depends only on the momentum
\begin{align}
    &2\hbar^{-1}f(E)~\ep = \Im\bar\Sigma^<=\Im\Sigma^<U^\dag = \Im ig^2G^<D^<U^\dag \nonumber\\
    &= \Im ig^2\sum_{n}\bar G^<_nD^< 
        \approx 2\hbar^{-1}f(E)\sum_{n:E_{n,p_z=0}<E}\ep_n
    \label{EqSigmaLL}
\end{align}
    and we arrive at \eqref{EqSigmaKin}.
    The sum runs over the LLs which cross the given energy level \(E\).

\subsection{Disorder}

In this section we consider electron self-energy due to scattering on impurities following \cite{Abramchuk2024}. 

The presence of impurities is modeled by electric potential $U(\vec{x}) = \sum_i u(\vec{x}-\vec{x}_i)$ with zero scattering length and uncorrelated distribution. 
The sum is over the positions of impurities located at $\vec{x}_i$. 
We employ the simplest possible model 
    $u(\vec{x}) = u_0 \delta(\vec{x})$. 
The correction to the electron Green function in the powers of \(U\) is given by
	\begin{align}
         \hat{G}_{xy}^{\text{{imp}}} &\approx  \hat{G}_{xy}  
            + \hbar^{-1}\braket{\int_z \hat{G}_{xz} \hat{U}_{\vec{z}}\hat{G}_{zy}} +\nonumber\\
         &+\hbar^{-2}\braket{\iint_{zz'} \hat{G}_{xz} \hat{U}_{\vec{z}}
            \hat{G}_{zz'} \hat{U}_{\vec{z'}}\hat{G}_{z'y}}+ \ldots  \label{EqGG} 
	\end{align}
with \(\hat{U} = {1 \otimes} U(\vec{x})\) in the triangle representation, which we adopted.
The propagator is to be averaged \(\braket{\ldots}\) over the disorder 
    --- the random positions of impurities.

The first order contribution can be absorbed by a redefinition of the Fermi level. 
The non-trivial leading effect of disorder is provided by the quadratic term.
Averaging over the disorder of the quadratic term of the expansion in \eqref{EqGG} reads
\begin{eqnarray}
    \hat{G}^{(2)}_\text{{imp}} &= &  \hbar^{-2}u^2_0 
        \langle \iint_{zz'} \hat{G}_{x z} \sum_{i}\delta(\vec z-\vec z_i) \hat{G}_{z z'}
    \times\\ && \times\sum_j \delta(\vec z'-\vec z_j) \hat{G}_{z' y} \rangle \nonumber\\ 
    & = & \hbar^{-2}u^2_0  \iint_{zz'} \hat{G}_{x z} n^{(2)}_\text{{imp}}(z-z') \hat{G}_{z z'} \hat{G}_{z' y},	\nonumber
\end{eqnarray}
where the pair correlator of the impurities potential is
\begin{equation}
	n^{(2)}_\text{{imp}}(z-z') \equiv \langle \sum_i \delta(\vec{z}-\vec{z}_i)   \sum_j \delta(\vec{z}'-\vec{z}_j) \rangle
\end{equation} 
where the averaging is over the different configurations of the coordinate sets $\{ ..., \vec{z}_1, \vec{z}_2, ... \}$. 
Assuming that the disorder correlation length is much smaller than $l_\mu$ 
    (the length scale defined by the Fermi level),
    we approximate
\begin{equation}
    n^{(2)}_\text{{imp}}(\vec z-\vec z') \approx n_\text{{imp}} \delta^{(3)} (\vec z-\vec z')
\end{equation}
where the delta-function is three-dimensional, so the correlator is time-independent.

The leading contribution by the disorder to the self-energy is
\begin{gather}
    \Sigma^R_\text{{imp}}(x,y) = \hbar^{-1}u^2_0 n_\text{{imp}} G^R({x, y}) \delta^{(3)}(\vec{x} - \vec{y}) 
\end{gather}
The space-time Fourier transform yields
    \(\delta(\vec x) \to 2\pi\delta(p_0)\),
    which means that scattering with any momentum transfer is equally possible
\begin{align}
    \Sigma^R_\text{{imp}}(p_0, \vec p) &= u_0^2n_\text{{imp}}\int \frac{d^3k}{(2\pi)^3}G^R(p_0, \vec p-\vec k) \label{EqSigmaImpP0}\\
    &\equiv \Sigma^R_\text{{imp}}(p_0) = {-}u_0^2n_\text{{imp}}\int \frac{d^3k}{(2\pi)^3}G^R(p_0, \vec k) \nonumber
\end{align}
In this approximation disorder can't generate contributions proportional to the spatial momentum.

\subsection{Acoustic phonons}

In this section we suggest a simplistic model for the electron-phonon interaction 
    limited by acoustic phonons,
    and calculate the corresponding electron self-energy.

Acoustic phonons are the modes of the density wave of the crystal lattice.
The interaction Hamiltonian reads (see e.g. Part 2, \S64 of \cite{LL9})
\begin{align}
    \hat H_\text{e-ph} &= \int d^{3}x ~\frac{w\hat\rho'}{\rho_0} 
        ~\hat\psi^\dag\hat\psi \label{EqHeph},
\end{align}
where \(w\) is the energy of volume deformation of the crystal lattice
(is of dimension of energy), 
\(\rho_0\) the unperturbed lattice density,
and \(\rho'\) is the density variation.

The interaction preserves chiral symmetry yet breaks the emergent relativistic invariance.

{In contrast to relativistic QFTs, where the speed of light is the same for fermions and gauge bosons, 
in  topological semimetals 
    the Fermi velocity \(v_F\) 
    (in topological semimetals as well as in graphene is of order \(10^{-2\div 3}c\) \cite{CMEZrTe5,Crassee2018})}
    is few orders of magnitude large then the sound velocity \(v_s\) 
    (which is of order of km/s as in any solid or liquid).
In particular, in \(Cd_3As_2\) \(v_F\sim 10^5~ m/s, 
~ v_s\sim 10^3~ m/s, 
~ u = v_s/v_F \sim 10^{-2\div 3}\) 
\cite{Crassee2018,Neupane_2014, Liang2022ultrafast, Wang2007computation}.

\(u\ll 1\) simplifies the NDT in a way that 
    the distribution-dependent terms in the Keldysh-Greens' functions 
    are time-independent,
    if we disregard wherever possible the terms proportional to \(u\) 
(see Problems to \S 92 in \cite{LL10} for the acoustic phonons Green functions)
\begin{align}
    D^R = D^{--}-D^{-+} = ~&\frac{\rho_0\bk^2}{(k_0+i0)^2-u^2\bk^2},\label{EqDss}\\
    D^< = D^{-+} = -i\pi\frac{\rho_0 |\bk|}{u}(&N_\bk\delta(k_0-u|\bk|) +\nonumber\\
    &+ (1+N_{-\bk})\delta(k_0+u|\bk|)) \nonumber\\
    \approx -i\pi\frac{\rho_0|\bk|}{u}&\coth\frac{\beta u|\bk|}{2}\delta(k_0),
\end{align}
where the initial distribution of phonons is thermal
\( 
    N_\bk = (e^{\beta uk}-1)^{-1} .
\) 
\rev{\textit{A posteriori}, 
    the approximate treatment of the phonons as quasi-static scatterers is self-consistent in either weak and strong magnetic field limits.}

Because of the relatively small sound velocity \(u\ll 1\), 
    the effective temperature for phonons \(\sim T/u\) is much larger than for electrons.
Thus we consider the thermal distribution for phonons 
    while the degenerate distribution for electrons,
    as in the previous section.

\subsection{Level width from electron Self-energy without magnetic field}\label{SectEpsB}

With \eqref{EqHeph}, \eqref{EqDss} we estimate the imaginary part of the self-energy operator. 
The simplest way to obtain the half-width is to calculate the imaginary part of the `lesser' component of the self-energy.
With the exact R/A Green functions for the Dirac fermion in magnetic field \cite{gorbar2014chiral},
at the Fermi level \(p_0=\mu\)
\begin{align}
    \Im&\frac\hbar4\tr\bar\Sigma^< =
    \Im\frac14\tr i\frac{w^2}{\rho_0^2}\int(d^4k) \times \nonumber\\
    &\quad\times 2\pi\hbar^{-1} i\delta((p-k)^2)(\slashed p - \slashed k )\gamma_0 f(p_0-k_0)\times\nonumber\\
    &\quad\times (-i)\pi\frac{\rho_0|\bk|}{u}\coth\frac{\beta u|\bk|}{2}\delta(k_0) \nonumber\\
    &\approx \frac{w^2}{\rho_0u} \frac{\mu f(\mu)}{8\pi^2\rev{\hbar^2}}
        \int d^3k\delta(\mu^2-({v_F}\bp-\bk)^2) |\bk|\coth\frac{\beta u|\bk|}{2} \nonumber\\
    &=  \frac{w^2}{\rho_0\rev{v_F^3}u}\frac{\mu f(\mu)}{4\pi\rev{\hbar^2}}\int_0^\pi d\Theta \sin\Theta\sin\frac\Theta2 
        \coth\left(\beta u\mu\sin\frac\Theta2\right) \nonumber\\
    &= 2f(\mu)\frac{w^2}{\rho_0\rev{v_F^3}u}\frac{\mu^3}{\pi\rev{\hbar^2} }\left(\frac{T}{u\mu}\right)^3
        \int_0^{\beta u\mu}dx ~x^2\coth x \nonumber\\
    &\sim 2f(\mu)\frac{1}{2\pi\rev{\hbar^2}}\frac{w^2}{\rho_0\rev{v_F^3}}\frac{\mu^2T}{u^2}, \quad \frac{u\mu}{T}\ll 1.
\end{align}
From the latter expression 
    we obtain the half-width
\begin{equation}
    \ep_\mu^\text{ph} = \frac{1}{{2}\pi\rev{\hbar^3}}{\frac{w^2}{\rev{v_F^3}\rho_0}\frac{\mu^2T}{u^2}} 
,
        \quad \frac{u\mu}{T}\ll 1.
    \label{EqEpsBw}
\end{equation}
For the disorder 
    calculation is simpler, see also \cite{Abramchuk2024}, 
    since the \(x\coth x\) part is replaced by a constant
\begin{equation}
    \ep_\mu^\text{imp} = \frac{\mu^2}{{2}\pi\rev{\hbar^3} v_F^2}u_0^2n_\text{imp}
.
\end{equation}
The total half-width then reads
\begin{gather}
    \ep_\mu = \frac{e^2\mu^2}{2\pi^2\rev{\hbar^2v_F}}(\rho_\text{ph}'T+\rho_\text{imp}), \label{EqEpsMu} \\
    \rho'_\text{ph} = \frac{\pi}{N_f\hbar e^2v_F^2}\frac{w^2}{\rho_0u^2}, \quad
        \rho_\text{imp} = \frac{\pi}{N_f\hbar e^2v_F^2}v_Fu_0^2n_\text{imp} \label{EqRhos}
\end{gather}
where we took into account the number of fermion flavors \(N_f\).

\subsection{Level width from electron Self-energy for Landau Levels}\label{SectEpsLLs}

As in the previous subsection, let us start with phonons only contribution.
For a given Landau level \(n\ge 0\), with magnetic length \(l\),
    and the spectral density given by Eq.(12) of  \cite{gorbar2014chiral}
\begin{align}
    \Im\frac\hbar4\tr\bar\Sigma^{<n}_\text{{ph}}&(p) = \Im\frac14\tr ig^2\int_k \nonumber\\
    &(\bar G^A-\bar G^R)\Big|_{p-k}f(p_0-k_0) D^<(k_0) \label{EqSigmaIntK3}\\
    =& \frac{w^2}{\rho_0^2}\int_k \pi\frac{\rho_0|\bk|}{u}\coth\frac{\beta u|\bk|}{2}\delta(k_0)\nonumber\\
    \times 2\pi\hbar^{-1}\delta((&p_0-k_0)^2-E_n^2(p_3-v_F^{-1}k_3))(p_0-k_0)f(p_0-k_0) \nonumber \\
    &\times \frac{(-1)^n}{2} e^{-(p-k)^2_\bot l^2}\Big(L_n-L_{n-1}\Big)\Big|_{2(p-k)^2_\bot l^2} \nonumber\\
    =& \frac{\pi w^2}{2\rho_0\hbar v_F^2u}p_0 f(p_0) \int_{k_3}\delta(p_0^2-(v_Fk_3)^2-2n\ep_L^2) \nonumber\\
    \times\int_{k_\bot} (-1)^n& e^{-k^2_\bot l^2}(L_n-L_{n-1})\Big|_{2k^2_\bot l^2}
        |\bk-\bp|\coth\frac{\beta u|\bk-\bp|}{2} \nonumber
\end{align}
Assuming the magnetic length large, 
    so that the transverse momentum \(k_\bot\) integral converges quickly,
    and with \( \frac{u}{T}\left(\frac{\hbar \rev{v_F^2}|e B|}{2\pi }\right)^{1/2} < \frac{u\mu}{T}\ll 1 \)
    we approximate 
\[|\bk-\bp|\coth\frac{\beta u|\bk-\bp|}{2} \approx \frac{2T}{u}.\]

Then to calculate the \(k_\bot\) integral we use some properties and power series definition of the Laguerre polynomials 
\begin{gather}
    L_n(x) - L_{n-1}(x) = \frac xn L'_n(x), \\
    L'_n(x) = -L^{(1)}_{n-1}(x), \quad
    L^{(\al)}_{n}(x) = \sum_{k=0}^n (-1)^k C_{n+\al}^{n-k}\frac{x^k}{k!} \nonumber
\end{gather}
where \(C_{n+\al}^{n-k} = \begin{pmatrix}n+\al\\n-k\end{pmatrix}\) are the binomial coefficients. 
Than the integral reduces to the Gamma function definition 
    \(\Gamma(z) = \int_0^\infty t^{z-1}e^{-t}dt , ~\Gamma(n) = (n-1)!\)
    and the sum of the binom
\begin{align}
    \pi^{-1}&\int d^2x_\bot e^{-x^2_\bot/2}(-1)^n(L_n(x_\bot^2)-L_{n-1}(x_\bot^2)) \nonumber\\
    & = \int_0^\infty dx e^{-x/2}(-1)^{n+1} (L_n(x)-L_{n-1}(x))\nonumber \\
    & = \int_0^\infty dx e^{-x/2}(-1)^{n+1}\frac{-x}{n}\sum_{k=0}^{n-1} (-1)^k C_{n}^{n-1-k}\frac{x^k}{k!} \nonumber\\
    & = \frac{(-1)^{n+1}}{n}\sum_{k=0}^{n-1}\frac{(-1)^{k+1}}{k!}C_n^{n-1-k}2^{k+2}(k+1)! \nonumber\\
    & = 4(-1)^{n}\sum_{k=0}^{n-1}(-2)^{k}\frac{(n-1)!}{(n-1-k)!k!} \label{EqSumPtn}\\
    & = 4(-1)^{n}\sum_{k=0}^{n-1}(-2)^{k}C^k_{n-1} = 4(-1)^{n+1}(1-2)^{n-1} = 4\nonumber .
\end{align}
while for  \(n=0, ~L_0=1, ~L_{-1}=0\) and 
\begin{gather}
    \pi^{-1}\int d^2x_\bot e^{-x^2_\bot/2} = 2.
\end{gather}

The half-width follows from temporal imaginary part of self-energy.
For LLL 
\begin{gather}
    \Im\frac\hbar4\tr\bar\Sigma^{<0}_\text{{ph}}(p) \approx 2 f(p_0) 
        \frac{|e B| w^2 T}{ {4}\pi u^2  \rho_0 }
\end{gather}
The factor of \(\mu^2\) in \eqref{EqEpsBw} is replaced here by the factor of \(B/2\),
    which corresponds to the densities of states,
    \(\sim\frac{4\pi\mu^2}{(2\pi)^2}\) on the FS and \(\sim\frac{B}{2\pi}\) on the LLL.
As expected, the function depends only on the energy component of the four-momentum.

In strong enough magnetic field, 
    if LL splitting is larger than the Fermi energy, 
    we recover the result of \cite{Abramchuk2024},
    with the disorder contribution added and with the notation of \eqref{EqRhos}
\begin{gather}
    \ep_\text{LLL} = \frac{|e^3B|\hbar v_F^2}{4\pi^2}(\rho_\text{ph}'T+\rho_\text{imp}), \label{EqEpsLLL} 
\end{gather}

For the higher Landau Levels the integral over \(k_3\) in \eqref{EqSigmaIntK3} 
    yields the square root in the denominator,
    and the half widths read 
\begin{gather}
    \ep_{n} = \frac{2\ep_\text{LLL}}{\sqrt{1-\frac{n}{\nu_B}}}, ~1\le n\le N_B,
    \quad \frac{u\mu}{T}\ll 1, 
        \label{EqEpsHLL} 
\end{gather}
where the number of HLLs that cross the Fermi level is 
\begin{gather}
    N_B = \left\lfloor \nu_B \right\rfloor, ~\nu_B=\frac{\mu^2}{2|eB|\hbar v_F^2} \label{EqNB},
\end{gather} 
So we find that the seemingly reasonable assumption that 
    all the higher Landau levels widths are approximately equal 
    is incorrect.
The half widths grow with LL number \(n\) approaching the Fermi surface.

The total half width (for given \(\mu\) and \(T\)), 
    which defines the magnetoconductivity, reads
\begin{align}
    \ep(B) &= \sum_{n=0}^{N_B}\ep_n \label{EqEpsMuB} =\\
    &= \frac{|e^3B|\hbar v_F^2}{{4}\pi^2}
        (\rho_\text{ph}'T+\rho_\text{imp})
        \sum_{n=0}^{N_B}\frac{(2-\delta_{n0})}{\sqrt{1-\frac{n}{\nu_B}}}
        \nonumber
\end{align}

The sum over a large number of LLs half widths converges  
    to the Fermi sphere half width \eqref{EqEpsMu}
\begin{gather}
    \ep(B)\Big|_{N_B\gg 1} = \sum_{n=0}^{N_B}\ep_n \approx \int_0^{\nu_B}dn ~\ep_n 
        = \ep_\mu
\end{gather}

\section{Comparison to Experiment}
\label{SectExp}

Let us suggest numerical values of model parameters.
On the experimental side the parameters are hard to measure and separate,
    including the LL splitting,
    but we find the numerical estimates important for assessing consistency of our model.
Thus, the numerical estimates are illustrative rather than quantitative.

In general, several lengths are to be compared, including  
    the mean free path length $l_0$, 
    the magnetic length $l_B = \sqrt{\frac{\hbar}{|e B|}}$, 
    the thermal length $l_T = \frac{\hbar v_F}{k_B T} $, 
    and $l_\mu = \frac{\hbar v_F}{\mu}$.

Throughout the text we use the example of $Cd_3As_2$ as a reference in order to evaluate the order of magnitude of the obtained physical values. 
For example, for Cd$_3$As$_2$ 
    with Fermi velocity $v_F \sim c/200$, 
    the mean free path may reach $l_0 \sim 100 \mu$m, 
    and at $T = 10 $K, $\mu \sim 100$ meV and $B = 1 $T we have 
    $l_B \sim 6 \times 10^{-8}m \sim l_\mu \sim 6\times 10^{-8}m \ll 10^{-5}m = l_T \ll l_0 \sim 10^{-4}m $, 
The magnetic scale and the scale of the chemical potential are of the same orders of magnitude, 
    are much larger than the scales of mean free path and temperature. 

We will also use for the same purpose the case of \(ZrTe_5\), 
    in which the magnetoconductivity was originally observed. 
For the other Weyl and Dirac semimetals such estimates may be accepted as rough estimates.  
The strong magnetic field limit corresponds to 
\begin{equation}
    {|\mu|} \lesssim v_F \sqrt{2| e B| \hbar},   \label{EqApprMuB}
\end{equation}
that is $l_B \lesssim \sqrt{2} l_\mu$. 
Provided that also $l_B \lesssim l_T$,
    only the lowest Landau level contributes  the physical observables.

In a real experimental setup, as for ZrTe\(_5\) in \cite{CMEZrTe5}, 
    only temperature, magnetic field value and orientation can be controlled 
    --- all the other values are properties of a given crystal, 
    which might have abrupt temperature dependence outside of the region \(T\sim {30}\div70\) K.

For the Dirac semimetal  $ZrTe_5$ \cite{CMEZrTe5} we have 
    $\mu \sim 100 $meV, $v_F \sim c/300$.
At  $ B \ge 5 $T then 
    $ l_\mu \approx 4 \times 10^{-8}{~m} > l_B \approx {3} \times 10^{-8}$ m.
The strong magnetic field limit might be achieved already at $B=5$ T.  

In general, the magnetic field in our equations,
    which is the field inside the material,
    differs from the field outside because of the crystal's permeability \(\mu_B\).
The permeability introduces 
some additional, generally unknown, dependence on magnetic field and temperature, 
    \(B\to B\mu_B^{-1}(T,B)\).
The permeability is likely to be independent of the external magnetic field \(B\) in the weak field limit and to have a value close to unity, 
    but there is no reason to expect it to be independent of temperature.
In the data analysis we disregard this additional complication.

\begin{figure}
	\includegraphics[width=0.9\linewidth]{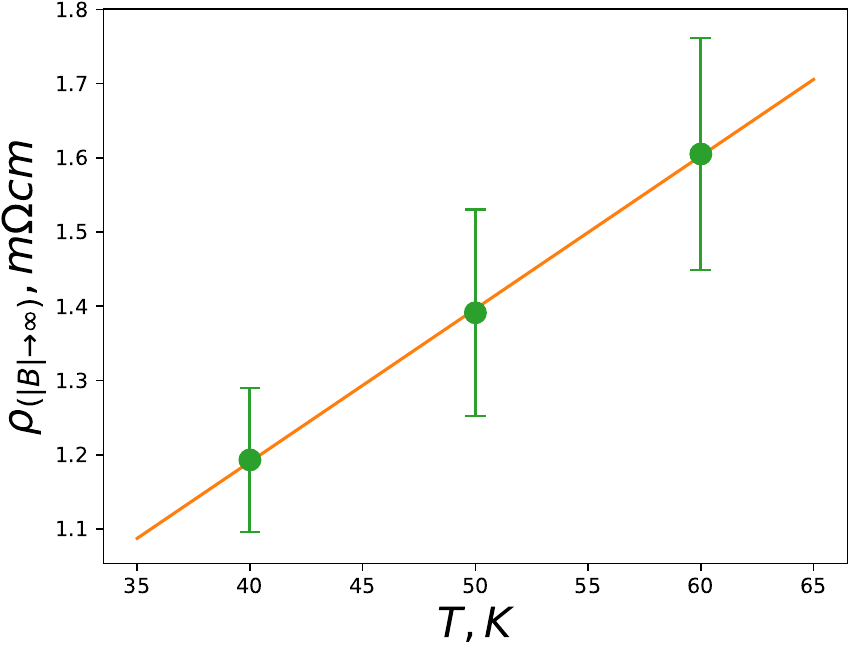}
	\caption{Large magnetic field resistivity of \(ZrTe_5\).
		Data extracted (points) from Fig.S1 of \cite{CMEZrTe5} 
		(see the {\sl Methods -- Transport Measurements} section in the latest arXiv version, 
            or Supplemental Materials in the journal version)
        {as the average at \(B>6\) T and \(B<-6\) T,}
		fitted with the obtained here dependence \eqref{EqPlCondLLL}
        \(\rho_{zz} = \sigma_{zz}^{-1}=\rho_\text{{imp}}+\rho_\text{ph}'T\);
        fitting parameters values from the data are \(\rho_\text{{imp}}\sim0.37\pm0.10~m\Omega cm\), \(\rho_\text{ph}'\sim0.021~m\Omega cm K^{-1}\){, which roughly correspond to  values of our model parameters} $w\sim {1.3}$ eV, $u_0 \sim {30}$ meV nm$^3$, $n_{imp} \sim   10^{-3}$ nm$^{-3}$. 
        The relatively large error bars result from asymmetry in the data at `positive' and `negative' magnetic fields, 
        while for `positive' (`negative') only data the errors are much smaller.
        Thus, the slope coefficient variance is small.
		\label{FigRhoLLLT}
	}
\end{figure}

From \cite{Abramchuk2024} we have \eqref{EqCondLLL} matching the data from Fig.S1 \cite{CMEZrTe5}
    as presented in Fig.\ref{FigRhoLLLT} 
\begin{gather}
    \sigma_\text{LLL}^{zz} =  \label{EqPlCondLLL}
        \sigma_\text{LLL} = \rho_\text{LLL}^{-1} = (\rho_\text{imp}+\rho'_\text{ph}T)^{-1}, \\
    \rho_\text{{imp}}\sim0.37\pm0.10~m\Omega cm,\quad\rho_\text{ph}'\sim0.021~m\Omega cm K^{-1},
\end{gather}
which roughly correspond to  values of our model parameters 
    $w\sim1.3$ eV, $u_0\sim30$ meV nm$^3$, $n_{imp}\sim10^{-3}$ nm$^{-3}$. 

\begin{figure}
	\includegraphics[width=0.9\linewidth]{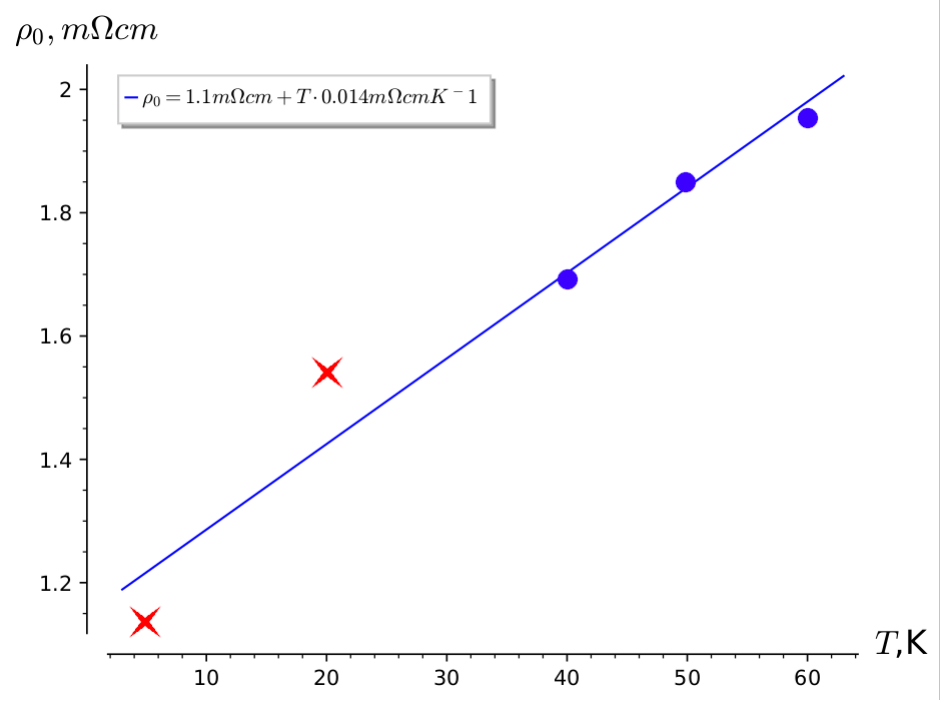}
	\caption{Zero magnetic field resistivity of ZrTe\(_5\).
		Data extracted (points) from Fig.2 of \cite{CMEZrTe5} 
		fitted with the obtained here dependence \eqref{EqPlCond0},
        fitting parameters values from the data are 
        \(\rho_{0,\text{imp}} \sim 1.1~m\Omega cm \sim 3\rho_\text{imp}\) and 
        \(\rho'_{0,\text{ph}} \sim 0.014~m\Omega cm K^{-1} \not\sim 3\rho'_\text{ph}\). 
        We omit error bars since we consider the analysis to be an estimation.
        The two low-temperature points are ill-defined 
            because of the cusp feature.
		\label{FigRho0T}
	}
\end{figure}

Our analytic result \eqref{EqCond0} suggests that the conductivities without magnetic field and in the strong magnetic field are proportional, because of the dimensional reduction factor
\begin{gather}
    \sigma_0^{jk}=\delta^{jk}\sigma_0,  
    \quad \sigma_0= \frac13\sigma_\text{LLL} = (\rho_{0,\text{imp}}+\rho'_{0,\text{ph}}T)^{-1}, \label{EqPlCond0} \\
    \quad\rho_{0,\text{imp}} = 3\rho_{\text{imp}},
    \,\rho'_{0,\text{ph}} = 3\rho'_{\text{ph}}. \nonumber
\end{gather}
with the same coefficients.
As presented in Fig.\ref{FigRho0T}, with the fitting parameters 
\begin{gather}
    \rho_{0,\text{imp}} \sim 1.1~m\Omega cm, \quad\rho'_{0,\text{ph}} \sim 0.014~m\Omega cm K^{-1},
\end{gather}
    the contribution of impurities matches the data Fig.2 of \cite{CMEZrTe5} within the uncertainties ,
    while the contribution of phonons fails to match even qualitatively  
    \(\frac{\rho'_{0,\text{ph}}}{\rho'_{\text{ph}}}\sim\frac12\) instead of 3.

The reason might be that the strong magnetic field is affecting motion of the ions in the crystal lattice,
    and thus modifies the dynamics of phonons, which we disregarded in our analysis,
    while for impurities this problem doesn't exist at all.

\begin{figure}
	\center{\includegraphics[width=0.9\linewidth]{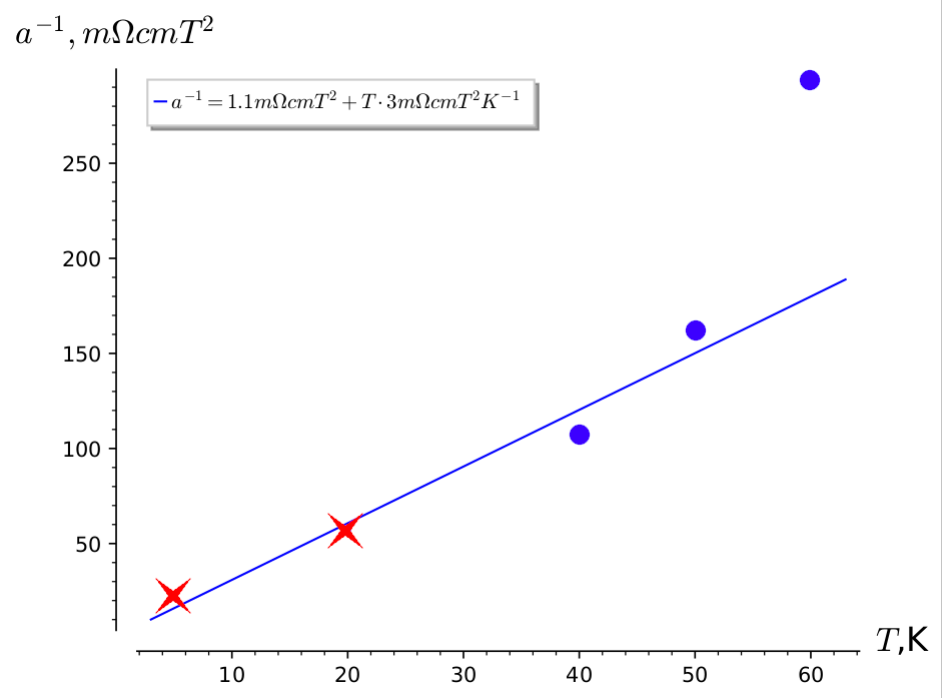}}
	\caption{Weak magnetic field magnetoconductivity of \(ZrTe_5\).
        Data extracted (points) from Fig.S1 (c.f. Fig.2) of \cite{CMEZrTe5}.
        If all points are included, the fit parameter value may become unphysical \(\al_\text{imp}<0\).
        The uncertainty is too large to derive parameters values.
        We omit error bars since we consider the analysis to be an estimation.
        The two low-temperature points are ill-defined
            because of the cusp feature.
        \label{FigaT}
	}
\end{figure}

With the notation of Fig.2 caption of \cite{CMEZrTe5}, 
    conductivity along the magnetic field is 
\begin{gather}
    \sigma^{(2)}_{zz}(B,T) = \sigma^{(0)}_{zz}(T) + B^2a(T),
\end{gather}
our analysis \eqref{EqCondWF}. \eqref{EqCond0} suggests 
\begin{gather}
    a^{-1}(T) = \al_\text{imp} + \al'_\text{ph} T, \label{EqaT}\\
    \al_\text{imp} = \frac{2\mu^4}{e^2\hbar^2v_F^4}\rho_{0,\text{imp}}, 
        \quad \al'_\text{ph} = \frac{2\mu^4}{e^2\hbar^2v_F^4}\rho'_{0,\text{ph}},\\
    \frac{2\mu^4}{v_F^4e^2\hbar^2} = 2\frac{l_{B}^4}{l_\mu^4}B^{-2}\Big|_{B=1T} \sim 10 \text{ T}^{-2}
    \nonumber
\end{gather}
As presented in Fig.\ref{FigaT},
the uncertainty is too large to derive parameters values,
    though the result might be consistent with the values for 
    \(\rho_{0,\text{imp}},\rho'_{0,\text{ph}}\) and 
    \(\rho_{\text{imp}},\rho'_{\text{ph}}\).

The discrepancies signalize problems in our model.


Apparently, our model for phonons is too simplistic for a quantitative description. 
Also, the formula \eqref{EqCondWF} might fail to reproduce \eqref{EqCondG}.
The data extraction and analysis procedures are rough estimates.

\section{Conclusions and discussion}\label{SectCon}

Starting with NDT for a semi-realistic model of a Dirac semimetal ---
    Dirac fermions in external electro-magnetic field \eqref{EqRelQ},
    that interact with disorder \eqref{EqSigmaImpP0} 
    and thermal ensemble of acoustic phonons \eqref{EqHeph},
we obtained kinetic equations in the relaxation time approximation \eqref{EqLLKin} 
    generally following \cite{Lin2019}.
To estimate the relaxation time, 
    parametrized with the level half-width \(\tau\approx\frac{\hbar}{2\ep}\) \eqref{EqEpsMuB},
    we calculated the imaginary part \(\ep\) of election self-energy with one-loop NDT. 
Then we solved the kinetic equation to obtain a new formula for longitudinal DC magneto-conductivity \eqref{EqCondIIExt},
    and suggested its relation to the experimentally observed resistivity \eqref{EqCondG}, 
    see also Fig.\ref{FigRhoG}.
In the weak \eqref{EqCondWF} and strong \eqref{EqCondLLLt}field limits our formula reproduces the CME-based patterns of magneto conductivity, 
    however we also specify the relaxation time, 
    which generally depends on magnetic field itself.

At small enough temperatures (few K), the non-monotonous magneto-conductivity is unexplained,
    and our approximation \eqref{EqEpsMu} might break down.
At large temperatures (dozens K) the model breaks down, 
    i.e. in ZrTe$_5$ because the Lifshitz transition \cite{CMEZrTe5} occurs \mbox{at \(T\sim 70\) K.}

From comparison to the experimental data (see Section \ref{SectExp}) for ZrTe$_5$ 
    we conclude that generally the model fails to provide quantitative description for 
    ZeTe$_5$ DC magneto-conductivity as function of temperature (up to 70 K) and magnetic field (up to 10 T).
The model for disorder might be satisfactory,
    while the model for phonons 
    be satisfactory only in the strong magnetic field limit.

An interesting qualitative conclusion regarding magneto-conductivity in topological materials 
    may be that
the notion of conductivity for an individual Landau levels is undefined, 
    since the collision integral scrambles electron states at all the Landau levels 
    that cross the Fermi level.
The scrambling is uniform because the disorder as well as phonons can absorb large momentum.

The lack of Landau quantization artifacts in experimental data on magneto-conductivity 
    in very strong magnetic field 
    compelled us to consider the actual observation process,
    to suggest and justify the averaging procedure \eqref{EqCondG}.

The latter subtlety suggests additional angle on the open problem of low temperature 
    non-monotonic dependence of conductivity on magnetic field.
Apart from the ideas that the behavior is dictated by magnetic properties of the crystal lattice,
    or that the behavior can be explained by our model 
    if we would not rely on the approximation listed in \eqref{EqEpsLLL},
    the behavior may be also rooted in the intricacy of the actual observation process
    (including but not limited by the misalignment problem discussed in \cite{CMEZrTe5}).

Another challenging goal could be to construct more realistic model 
    for particular Dirac and Weyl materials and fit experimental data for sets of observables.

In the general context of QFT, this paper exemplifies 
    derivation and application of a kinetic theory 
    derived from a model Hamiltonian 
    (without free `phenomenological' parameters as opposed to `model' parameters)
    using NDT (the Keldysh technique)
    in a general non-equilibrium setting,
    with additional symmetries, in non-conventional basis.

\section{Acknowledgments}

I am grateful to Prof. M.A.Zubkov for useful discussions and critical remarks.

\bibliographystyle{elsarticle-num-names}

\end{document}